# DISTANCE BASED ASYNCHRONOUS RECOVERY APPROACH IN MOBILE COMPUTING ENVIRONMENT

Yogita Khatri

Department of Computer Engineering, JSSATE , Noida, India
yogitabaveja65@gmail.com

## ABSTRACT

*A mobile computing system is a distributed system in which at least one of the processes is mobile. They are constrained by lack of stable storage, low network bandwidth, mobility, frequent disconnection and limited battery life. Checkpointing is one of the commonly used techniques to provide fault tolerance in mobile computing environment. In order to suit the mobile environment a distance based recovery scheme is proposed which is based on checkpointing and message logging. After the system recovers from failures, only the failed processes rollback and restart from their respective recent checkpoints, independent of the others. The salient feature of this scheme is to reduce the transfer and recovery cost. While the mobile host moves with in a specific range, recovery information is not moved and thus only be transferred nearby if the mobile host moves out of certain range.*

## 1. INTRODUCTION

A distributed system consists of distinct processes which are geographically separated and communicate with each other by exchanging messages. A mobile computing system is a distributed system where some processes are running on Mobile Host (MH) that can move. To communicate with MHs, Mobile Support Stations (MSSs) are added. MSSs communicate with each other through a wired network. MHs communicate with MSSs on wireless network. Mobile computing system are constrained by lack of stable storage, low bandwidth, limited battery life, frequent disconnection .These unique attributes of mobile computing environment call for a different approach to checkpointing and recovery that are less demanding on the network bandwidth, computing power and storage capacity. Checkpointing consist of intermittently saving the state of the program in a reliable storage medium. Upon detection of a failure previous consistent state is restored. In case of a failure, checkpointing enables the execution of a program to be resumed from a previous consistent state rather than starting from the beginning. Checkpoint protocol are less restrictive and easier to implement.

The two fundamental approaches for checkpointing and recovery are (i) Asynchronous Checkpointing (ii) Synchronous Checkpointing. In asynchronous (uncoordinated checkpointing) checkpointing, a process can take a checkpointing on its own independent of the other process, periodically. Here no restriction is placed on the process on when to take the checkpoint. Each process can take as per its need. After recovery from a failure, a failed process rollback to its recent checkpoint and communicate with the dependent processes to build a consistent state. This scheme suffers from the domino effect due to the absence of coordination among processes which may cause the loss of large amount of useful work , possibly all the way back to the beginning. Many algorithms have been proposed to cope up with this effect which uses the concept of message logging [5], [6]. Uncoordinated checkpointing forces each process to maintain multiple checkpoints and to invoke a garbage collection algorithm periodically to reclaims the checkpoint that are no longer needed. In order to determine a consistent global state during recovery, the processes need to record the dependencies among their checkpoints during failure- free operation.

On the other hand, in synchronous (coordinated checkpointing) approach processes coordinate with each other to achieve a global set of consistent checkpoints. Due to coordination, it is free from domino effect. It requires each process to maintain only one permanent checkpoint on stable storage, reducing the storage overhead and eliminating the need for garbage collection.

This scheme assumes that processes needs to stop their computation while this checkpoint activity is going on.

### 1.1. Related Work

Application failure recovery in the mobile computing environment has received considerable attention in the recent years. The scheme that has been proposed in this paper employs checkpointing and message logging both.

Much of the previous work [7],[8],[9],[10] in synchronous checkpointing has focussed on minimizing the number of synchronous messages and the number of checkpoints during checkpointing. They are blocking algorithms, that is, processes need to stop their computation during synchronization activities which degrade the performance of mobile computing system.

Recently, nonblocking algorithms [11], [12] have gained considerable attention. In these algorithm processes need not to block during checkpointing by using a checkpoint sequence number to avoid inconsistencies. In mobile computing system, since checkpoints need to be transferred to the stable storage at the MSS over the wireless network, taking unnecessary checkpoints may waste a large amount of wireless bandwidth

The Prakash-Singhal algorithm [1] was the first algorithm to combine these two approaches. It .only forces a minimum number of processes to take checkpoints and does not block the underlying computation during the checkpointing.

Acharya et al. [2] describes uncoordinated checkpointing, where multiple MHs can arrive at a global consistent checkpoint without coordination messages. However, neither it takes into account how failure recovery is achieved nor does it address the issue of recovery information management in the face of MH movement.

P. Kumar and A. Khunteta [3] proposed a minimum-process coordinated checkpointing algorithm for deterministic mobile distributed systems, where no useless checkpoints are taken, no blocking of processes takes place, and anti-messages of very few messages are logged during checkpointing. In their algorithm they have tried to reduce the loss of checkpointing effort when any process fails to take its checkpoint in coordination with others.

The authors in [4] presents a low overhead recovery scheme based on a communication induced checkpointing, which allows the processes to take checkpoints asynchronously and uses communication-induced checkpoint coordination for the progression of the recovery line. The scheme also uses selective pessimistic message logging at the receiver to recover the lost messages. However, the recovery scheme can handle only a single failure at a time.

### 1.2. Problem Formulation

The objective of the present work is to design a mechanism for checkpointing and recovery in mobile computing environment which will ensure efficient use of limited resources. In this paper, emphasis is given on message logging and an efficient scheme is proposed to implement logging with a small overhead for the mobile computing environment.

To reduce the stable storage access cost and to cope with the space problem of mobile hosts (MH), the task of logging is assigned to the mobile support stations (MSSs). Since messages heading to the mobile hosts are routed through the mobile support stations, message logging by the mobile support stations does not impose any extra communication overhead. As a MH moves around the cells, the checkpoints and the message log of MH becomes scattered over various MSSs. In case of failure, MH must locate the latest checkpoint and also has to locate the sequence of logged messages, which in turns increases the recovery cost. For fast recovery, it is required that the checkpoints and message log must be nearby. Furthermore, if checkpoints and message log are transferred with every handoff, then the transfer cost will become very much significant. Thus in this paper, I proposed a distance based recovery scheme which restrict the transfer of recovery information as MH moves across the cells and is allowed only when it moves out of a particular range.

This paper is organised as follows. In section 2 we have stated system model. Section 3 is explaining the data structure needed to support the proposed technique. In section 4, we have explained the types of messages transferred across the system. Section 5 explained the checkpointing and message logging. In section 6, we have explained the procedure of asynchronous recovery. Finally section 7 draws conclusion.

## 2. SYSTEM MODEL

We are considering a mobile computing environment with a network consisting of stationary and mobile hosts. A mobile host can change its location and network connection while computations are being processed. Message passing between two hosts is enabled via. the mobile support stations(MSS). The MSSs are reliable and are interconnected by a wired network. A MSS handles all communication to and from MHs within its area of influence known as a cell usually determined by the range of wireless transmission. Each MSS has a fixed wireless transmission range known as a cell and an MH can move from one cell to another. Assuming a hexagonal shape for each cell, a hexagonal network coverage model will be formed by a community of cells. Thus sending a message to another MH consist of two one-hop wireless transmission between the sender and receiver MHs and their respective local MSS in addition to an arbitrary number of hops across the wired infrastructure between the sender's MSS and receiver MSS. At any time, a MH can be connected to at most one MSS. Channels are virtually lossless and they ensure FIFO communication and all communication takes place through messages. The interaction between the MH and the network infrastructure most relevant to failure and recovery are hand-offs, disconnect and reconnect.

Considering that the MH's disk cannot be assumed to be stable, each MSS is equipped with enough volume stable storage to store the state and log information for all the MHs currently in its cell as well as those that were recently in its cell. However due to the fact that MSSs must support multiple concurrent MHs, this storage must be efficiently managed.

## 3. NOTATIONS AND DATA STRUCTURES USED

We use the following notations in describing our approach. The $i^{th}$ mobile host is denoted by $h_i$ and the $p^{th}$ mobile support station is denoted by $S_p$. $M_i^a$ represent the $a^{th}$ application message generated by host $h_i$. $C_i^a$ represent the $a^{th}$ checkpoint taken by host $h_i$.

### 3.1. Data Structure Maintained at Each MH

Let $C_i^a$ be the $a^{th}$ checkpoint taken by host $h_i$. Two integers $i$, $a$ are used as the identifier for $C_i^a$. Each mobile host $h_i$ also maintains a counter $rec\_seq_i$, that denotes the number of messages $i^{th}$ mobile host has received.

## 3.2. Data Structure Maintained at Each MSS

Active_MH_list : Active_MH_list contains the list of MH identifiers that are supported in the cells associated to different MSSs.

Disconnect_MH_List : It contains the list of the MH , that are no more connected to the MSS.

Message log : MSS maintains a message log for each MH , $h_i$ , that contains the messages headed for that $h_i$ along with the sequence number of the message which will be stored in the variable $num\_msg_i$.

Trace Record : MSS maintains a trace record $Trace_i$ for each mobile host $h_i$ that consist of two integer variables cp_seq and cp_loc and a list called log_set. cp_seq denotes the checkpoint sequence number and the cp_loc identifies the MSS that stores that checkpoint. log_set includes the list of the MSSs which carry the message log for $h_i$.

## 4. TYPES OF MESSAGES

1) Application Messages: Application messages are those which are passed between the mobile hosts through mobile support stations during computation.

2) Mobility Based Messages: These messages are related to the mobility.

Leave(r): When a MH connected to a MSS leaves its current cell, it sends the leave(r) message to the MSS , where r denotes the sequence number of the last message received from the MSS.

Join(MH-id , previous MSS-id): It is sent by the MH to the new MSS indicating its id and the id of the MSS to which it was connected previously.

Disconnect(r): When a MH goes into sleep mode for power consumption, without disconnecting itself from the local MSS, it send the disconnect(r) message to the MSS, where r is the sequence number of the last message received. When MSS receives the disconnect message from the MH, it marks the MH as disconnected by setting a flag in the Disconnect_MH_List.

Reconnect (MH-id,previous-id) : When a MH wants to
reconnect to any MSS , it send the reconnect message ,where MH-id indicates its own id and previous MSS-id is the id of the MSS to which it was connected previously.

## 5. MESSAGE-LOGGING

Each mobile host $h_i$ periodically takes a checkpoint independent of others. It increments the checkpoints sequence number by one every time it takes a new checkpoint. After taking checkpoint , let say $C_i^a$ , it sends the checkpoint with its identifiers (i , a) to its current MSS say $S_p$. It also delivers the sequence number of the message it received last, prior taking the checkpoint. On receipt of the checkpoint, $S_p$ saves the checkpoint and the related information into the stable storage. Whenever a message $M_i^a$ comes for $h_i$ , $S_p$ first log the message in the message log, together with the sequence number 'a', and then delivered to $h_i$. $S_p$ also logs the messages related to the mobility such as join , leave, disconnect and reconnect , received from the MHs. Any of these messages sent from the mobile host $h_i$ , must carry the value of $rec\_seq_i$.

During the handoff procedure, that is, when a mobile host $h_i$ moves from the old MSS to new MSS , as shown in the fig 1, $Trace_i$ , which was maintained by old MSS get transferred to the new MSS. New MSS , then saves the $Trace_i$ record into its stable storage. Now when the new MSS saves a new checkpoint for the $h_i$, it puts the checkpoint sequence number in the $Trace_i.cp\_seq$ and its id into $Trace_i.cp\_loc$ and makes the $Trace_i.log\_set$ list empty, so that it can include only the MSSs which have saved the message logs after the latest checkpoint.

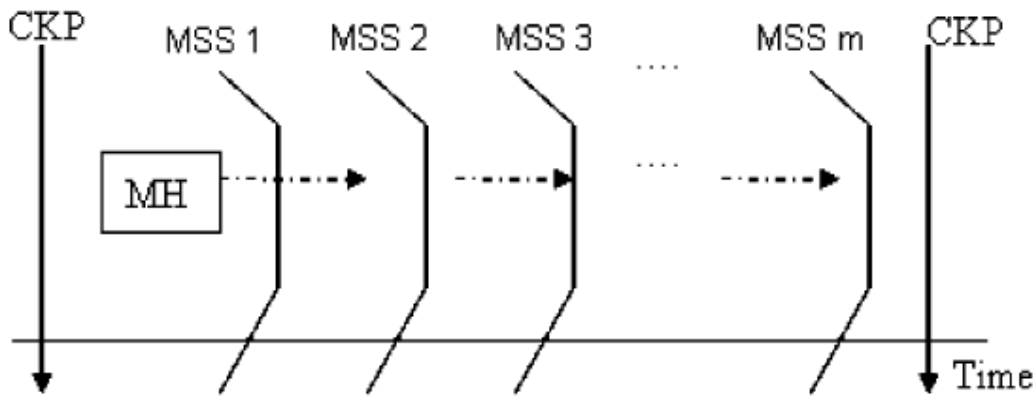

Fig 1. Checkpointing during hand-off

As a mobile host $h_i$ moves from one MSS to another, the message logs of $h_i$ becomes distributed over the stable storages at various MSSs it has visited. Now consider the situation that a failure has occurred at $h_i$ and its latest checkpoint lies at MSS which is quite far away, then in this case, the cost of obtaining the checkpoint and the corresponding message log will be very high. Therefore a distance based scheme for recovery is proposed in this paper. Idea is that, the checkpoint and message logs need to be moved into a new MSS during the handoff only when the moving distance of the MSS to which mobile host is connected from the MSS carrying the latest checkpoint exceeds a certain value K. Each MSS maintains a distance table that include the distance between any other MSS and itself. Let us consider $D_{i,p,q}$ is the distance between the mobile support station $S_p$ in which $h_i$ is currently residing and the mobile support station $S_q$ carrying the latest checkpoint. After each handoff, the new MSS, that is , $S_q$ calculates the distance $D_{i,p,q}$ and perform normal handoff if $D_{i,p,q} < K$. Only when $D_{i,p,q} > K$ , then the latest checkpoint and message logs are transferred to the new MSS $S_q$.

/* Message logging Mechanism for $h_i$ handled by $S_p$ */

Message logging( )
{
    When $S_p$ receives a message M from $h_i$
    If (M is in the list{join,leave,reconnect,disconnect})
      Save[ M, $rec\_seq_i$ ] into stable log space
    Else if ( M is an application message and want to
                deliver to $h_j$) then
      {
        $num\_msg_j = num\_msg_j + 1$
        Save[ M , j ,$num\_msg_j$] into stable storage
        $Trace_j.log\_set = Trace_j.log\_set \cup p$;
        Deliver M to $h_i$
      }
    During handoff when $h_i$ enters into a cell managed
    by $S_q$
        $S_q$ saves [ $Trace_i$, $rec\_seq_i$ ]into the stable storage.
}

/* Checkpointing and Updating of Trace record */

Let's consider $h_i$ is connected to MSS $S_q$, wants to take a new checkpoint.
Checkpoint( )
{
    $Chknum_i = chknum_i +1$;
    Save [i, $chknum_i$ , $rec\_seq_i$] with checkpoint
    Send [i, $chknum_i$ , $rec\_seq_i$] to $S_q$ along with checkpoint
    When $S_q$ receives a checkpoint from $h_i$, save it into the
    stable _checkpoint_space.
    $Trace_i.cp\_seq = chknum_i$;
    $Trace_i.cp\_loc = q$;
    $Trace_i.log\_set = null$;
}

/*Handoff Management*/

Handoff ( )
{    When $S_p$ receives a join message from $h_i$, it obtain the $Trace_i$ record from the old MSS $S_q$
      If ($D_{i,q,p} >= K$)
      {
      Obtain the latest checkpoint of $h_i$ from $S_q$ corresponding to the checkpoint sequence number
        residing in the $Trace_i.cp\_seq$ and also obtain all the messages $M_i^a$ where $a > rec\_seq_i$.

    /* obtain the message log of $h_i$ , which is distributed over various MSS */

      For (each MSS $S_r$ belongs to $Trace_i.log\_set$)
      Obtain all messages $M_i^a$ from $S_r$ such that ($a > rec\_seq_i$)
      Save the checkpoint and message log obtained in the stable storage
      $Trace_i.cp\_loc=p$;
      $Trace_i.log\_set=p$;
      }
      Else /* Normal handoff */
      {

        Save[ ( $Trace_i$ , $rec\_seq_i$)] into the stable storage
      }
}

## 6. ASYNCHRONOUS RECOVERY

When any of the mobile host get failed, all it has to do is, to just rollback to its latest checkpoint, independent of the others. No other mobile host need to roll back together. Whenever a mobile host recovers from a failure, it first of all obtain the latest checkpoint indicated by $Trace_i.cp\_seq$ from the MSS identified by $Trace_i.cp\_loc$ and then obtain all messages which are sent to the mobile host during its failure time , that is, after the checkpoint, from all the MSSs contained in $Trace_i.log\_set$.

## 7. CONCLUSIONS

In this paper, I have presented a distance based asynchronous recovery scheme based on checkpointing and message logging for mobile computing systems. The attention is given on reducing the transfer cost and recovery cost because as MH moves across the cells, its message log becomes distributed over a number of MSSs. If we transfer checkpoints and message log with every handoff, then the transfer cost will become very much significant and also puts extra overhead on the network bandwidth. Thus, in this paper we restrict the transfer of the recovery information as MH moves across the cells and it is allowed only when it moves out of a particular range.